\documentclass[12pt]{article}
\usepackage{amsfonts}
\usepackage{amsmath}
\usepackage{amssymb}
\usepackage[title,titletoc]{appendix}
\usepackage[english]{babel}
\usepackage{comment}
\usepackage{color}
\usepackage{geometry}
\usepackage{graphics}
\usepackage{graphicx}
\usepackage[colorlinks=true,urlcolor=blue,linkcolor=blue,citecolor=blue]{hyperref}
\usepackage{multicol}
\usepackage{multirow}
\usepackage{natbib}
\usepackage{times}

\addto\extrasenglish{}

\renewcommand{\baselinestretch}{1.5}

\newcommand{\vd}{\mathbf{d}}
\newcommand{\vn}{\mathbf{n}}
\newcommand{\vt}{\mathbf{t}}
\newcommand{\vx}{\mathbf{x}}
\newcommand{\vz}{\mathbf{z}}

\newcommand{\vmu}{\text{\boldmath{$\mu$}}}
\newcommand{\vsigma}{\text{\boldmath{$\sigma$}}}
\newcommand{\vpi}{\text{\boldmath{$\pi$}}}
\newcommand{\vtheta}{\text{\boldmath{$\theta$}}}

\newcommand{\mTheta}{\mathbf{\Theta}}

\newcommand{\N}{\mathcal{N}}
\newcommand{\B}{\mathcal{B}}
\newcommand{\D}{\mathcal{D}}
\newcommand{\G}{\mathcal{G}}
\newcommand{\LN}{\mathcal{LN}}
\newcommand{\PO}{\mathcal{PO}}

\title{
	\textbf{
		Bayesian Analysis of Mixtures of Lognormal Distribution with an Unknown Number of Components from Grouped Data
	}
	\thanks{
		Previous versions of this paper were presented at CFE 2019 in London as well as the seminars at Vienna University of Economics and Business, Keio University, Kobe University, and Kwansei Gakuin University.
		We would like to thank the seminar/conference participants, especially Duangkamon Chotikapanich for her valuable comments and suggestions.
		Part of this research was conducted while the author was visiting Institute for Economic Geography and GIScience, Vienna University of Economics and Business, whose hospitality is gratefully acknowledged.
		This work is partially supported by KAKENHI \#20H00080, \#20K01590 and \#16KK0081.
	}
}
\author{
	Kazuhiko Kakamu
	\thanks{
		School of Data Science, Nagoya City University, Yamanohata 1, Mizuho-cho, Mizuho-ku, Nagoya 467-8501, Japan.
		Email: \texttt{\href{mailto:kakamu@ds.nagoya-cu.ac.jp}{kakamu@ds.nagoya-cu.ac.jp}}
	}
}
\date{}
\begin{document}
\maketitle
\begin{abstract}
	This study proposes a reversible jump Markov chain Monte Carlo method for estimating parameters of lognormal distribution mixtures for income.
	Using simulated data examples, we examined the proposed algorithm's performance and the accuracy of posterior distributions of the Gini coefficients.
	Results suggest that the parameters were estimated accurately. Therefore, the posterior  distributions are close to the true distributions even when the different data generating process is accounted for. 
	Moreover, promising results for Gini coefficients encouraged us to apply our method to real data from Japan.
	The empirical examples indicate two subgroups in Japan (2020) and the Gini coefficients' integrity.
	\vspace*{8pt}

	\vspace*{8pt}

\end{abstract}

\clearpage
\section{Introduction}
Finding a distribution that fits the data well is one of the main challenges in the estimation of income distributions.
However, we face the trade-off between the interpretation of parameters and the fit of the hypothetical distribution.
To explore the fit of the distribution, several flexible hypothetical distributions are proposed, including: the generalized beta distribution of first and second kind \citep{M84}; generalized beta distribution \citep{MX95}; double Pareto-lognormal distribution \citep{RJ04}; and $\kappa$-generalized distribution \citep{CGK07}.
Several of these support interpretations that are economically meaningful.
\footnote{
	To better fit income distribution models to empirical data, Bayesian Model Averaging (BMA), as explored by \citep{GCR05}, has also been proposed as a viable approach.
}<++>

Conversely, the mixture distribution models are also considered to fit the distribution to the data because the assumed underlying distributions are easy to interpret and the distribution fits better than single component models in many cases.
The greater level of detail offered by mixture distribution models, such as a subgroups' information, is evident from the model's adoption in \citet{PvD98,GH12}, among other studies.

Mixture distribution models have also considered the framework of household income distributions from a Bayesian point of view using Markov chain Monte Carlo (MCMC) methods.
For example, in the case of lognormal distribution, \citet{LN16} considered a finite mixtures of lognormal (MLN) distribution model from individual data and determined the number of components by the marginal likelihood \citep{C95} and DIC \citep{SBCvdL02}.
The income inequality was then decomposed into between-subgroup and within-subgroup components.
Moreover, it is also considered in gamma distribution cases.
\citet{WIR01} considered the mixtures of gamma distribution model with a known and unknown number of components.
\citet{CG08} examined the Canadian income data using two components' mixtures of gamma densities, which is the known number of components case in \citet{WIR01}.
However, with the exception of \citet{WIR01}, the number of components were assumed in advance or determined after estimation in these studies, and they used individual or household data as mentioned above.

As with \citet{WIR01}, there are two main approaches for dealing with an unknown number of components in a mixture model: one uses a Dirichlet process prior \citep{EW95}, and the other uses a reversible jump MCMC algorithm \citep{RG97}, which is used in \citet{WIR01}.
The reversible jump MCMC algorithm, which was first proposed by \citet{G95}, is one of the most powerful tools in model determination.
\citet{RG97} proposed the algorithm in the framework of the mixtures of normal distribution model.
Subsequently, some scholars have proposed extensions to the multivariate normal distribution \citep{K09} and the mixtures of normal distribution with the same component means \citep{PI09}.
In addition, \citet{MH13} pointed out that the posterior from a Dirichlet process prior for the number of components was not consistent---unlike with the reversible jump MCMC, the Dirichlet process prior did not converge at the true number.
Therefore, we consider the reversible jump MCMC algorithm in this study, because we are also interested in the number of components in the analysis of income distribution.

Although the availability of individual and household data has improved, it remains difficult to access, especially in developing countries.
Alternatively, the grouped data, which partitions the sample space of observations into several non-overlapping groups, is widely available.
Using this type of data, \citet{GdL14} considered the MCMC sampling scheme for finite mixtures of normal distribution.

This study extends their approach in two significant directions.
First, we generalize the assumed distribution in \citet{GdL14} from the normal distribution to the lognormal distribution.
This allows for a more realistic modeling of income data, which is typically skewed and strictly positive.
Second, instead of fixing the number of components in advance, we adopt the reversible jump MCMC algorithm proposed by \citet{RG97}, enabling us to estimate the number of components directly from the data.
This extension enhances the model's flexibility and allows it to capture potential overdispersion in the structure of income subgroups, which is especially valuable when analyzing heterogeneous populations based on grouped data.

Exploring this model is worthwhile because the number of components provides information about population subgroups, as discussed in \citet{LN16}.
Therefore, if it is possible to determine the number of components from grouped data, this approach can be used for detailed comparisons of income inequalities in developing countries.

This study aims to develop a reversible jump MCMC method for the mixtures of lognormal (MLN) distribution model from grouped data to examine the income distributions and income inequalities in Japan.
Our proposed algorithm is discussed using simulated data examples.
From these, we can confirm that our proposed algorithm works well in terms of the accuracy of the parameters and in fitting the distribution.
The data also suggests that the posterior distributions of the Gini coefficients are accurate.
Hence, we applied it to real data in Japan in 2020 to examine the income distributions and inequalities.
From the results, we identified two subgroups in both two-or-more person households and workers' households.
We also observed that the Gini coefficient of two-or-more person households was larger than that of workers' households.

The rest of this paper is organized as follows.
In Section \ref{sec:LMM}, we summarize the MLN distribution model using grouped data with its Gini coefficient and obtain a joint posterior distribution.
Section \ref{sec:PA} discusses the computational strategy of the MCMC method.
In Section \ref{sec:NESD}, our approach is illustrated using simulated datasets.
Section \ref{sec:ARD}, examines the empirical examples using real datasets from Japan.
Finally, brief conclusions are offered in Section \ref{sec:C}.

\section{Mixtures of Lognormal Distribution Model using Grouped Data\label{sec:LMM}}
Let $x > 0$, which means the annual income of households or individuals, for example, follow any hypothetical distribution. Let $x_{i}$, $i = 1, 2, \ldots, n$ observations be sampled from the distribution.
Then, the grouped data partitions the sample space of observations into $K > 1$ non-overlapping intervals of the forms $(t_{0}, t_{1}]$, $(t_{1}, t_{2}]$, $\ldots$, $(t_{K-1}, t_{K})$, where $t_{0} = 0$ and $t_{K} =\infty$.
Moreover, only the number, $n_{k}$ of observations falling in each interval $(t_{k-1},t_{k}]$, $k = 1, 2,\ldots, K$, can be observed with $\displaystyle \sum_{k = 1}^{K} n_{k} = n$.
It should be mentioned that the class income mean $\bar{x}_{k}$, which means the average of $x_{i}$ in the interval $(t_{k-1},t_{k}]$, is also available in many cases.

Let $\vtheta$ be the vector of parameters of any underlying hypothetical distribution, which we assume in advance.
Let $f(x|\vtheta)$ and $F(x|\vtheta)$ be the probability density function (PDF) and cumulative distribution function (CDF), respectively.
Given the PDF and CDF, we define the likelihood function, which is based on the concept of selected order statistics, to estimate the parameters of the distribution.
\footnote{
	\citet{MR79} considered the likelihood based on the multinomial distribution, whereas \citet{NK11} considered the likelihood based on the selected order statistics.
	As is pointed out by \citet{EB21}, the likelihood based on the multinomial distribution is applicable to the data with known fixed boundaries and random frequencies, while the likelihood based on the selected order statistics is applicable to the data with known random boundaries and fixed frequencies.
	In this study, we follow the likelihood based on \citet{NK11}, because we are interested in the decile data, whose features are with known random boundaries and fixed frequencies.
	It should be mentioned that our approach merely treats the special case of DGP1 in \citet{EB21}.
}
To explain the likelihood function, let $\vt = (t_{1}, t_{2}, \ldots, t_{K-1})^{\prime}$ be the vector of the endpoints of the intervals and let $\vn = (n_{1}, n_{2}, \ldots, n_{K})^{\prime}$ be the vector of frequencies, which fall in the intervals.
Then, the likelihood function is defined as follows:
\begin{eqnarray}
	L(\vt | \vtheta, \vn) = n! \frac{F(t_{1}|\vtheta)^{n_{1} - 1}}{(n_{1} - 1)!} f(t_{1}|\vtheta) \left\{ \prod_{k = 2}^{K-1} \frac{\left( F(t_{k}|\vtheta) - F(t_{k - 1}|\vtheta) \right)^{n_{k} - 1}}{(n_{k} - 1)!} f(t_{k}|\vtheta) \right\} \frac{\left( 1 - F(t_{K-1}|\vtheta) \right)^{n_{K}}}{n _{K}!}.
	\label{eq:L}
\end{eqnarray}
Once the parameter estimate for $\vtheta$ is obtained from \eqref{eq:L} using maximum likelihood and so on, the Gini coefficient can be estimated by using
\begin{eqnarray}
	G = -1 +\frac{2}{\mu} \int_{0}^{\infty} x F(x|\vtheta) f(x|\vtheta) dx,
	\label{eqn:gini}
\end{eqnarray}
where $\mu$ is the mean of the distribution.
\footnote{
	In the numerical integration, we use the expression
	\begin{eqnarray*}
		G = 1 - \frac{\displaystyle \int_{0}^{\infty}(1 - F(x|\vtheta))^{2}dx}{\displaystyle \int_{0}^{\infty}(1 - F(x|\vtheta)) dx},
	\end{eqnarray*}
	because it is equivalent to \eqref{eqn:gini} \citep[see][]{D79} and easier than calculating \eqref{eqn:gini}.
}

In the empirical analysis we need to specify the hypothetical income distribution.
First, we start with the lognormal (LN) distribution, following \citet{NK11}, because the distribution fits to the Japanese data, which is also used in this empirical example.
Although we could consider the other distributions, such as a gamma distribution and so on, we restrict our discussion on the LN distribution to focus on our empirical example.
Let $x \sim \LN(\mu, \sigma^{2})$, which means $x$ follows LN distribution, where the PDF is expressed by
\begin{eqnarray}
	f(x|\mu,\sigma^{2}) = \frac{1}{\sqrt{2 \pi \sigma^{2}}x}\exp\left\{ -\frac{(\ln x - \mu)^{2}}{2\sigma^{2}} \right\},
	\label{eq:PDF-LN}
\end{eqnarray}
and the CDF is expressed by
\begin{eqnarray}
	F(x|\mu,\sigma^{2}) = \Phi\left( \frac{\ln x - \mu}{\sigma} \right),
	\label{eq:CDF-LN}
\end{eqnarray}
where $\Phi(\cdot)$ is the CDF of the standard normal distribution.
If we substitute \eqref{eq:PDF-LN} and \eqref{eq:CDF-LN} for \eqref{eq:L}, it becomes the likelihood function for the LN distribution model and its Gini coefficient has a closed form, expressed by
\begin{eqnarray}
	G_{LN} = 2\Phi\left( \frac{\sigma}{\sqrt{2}} \right) - 1.
	\label{eqn:gini-LN}
\end{eqnarray}

To extend the above results, we consider the MLN distribution model with $R$ components.
Let us begin with the fixed number of components model.
Let $\vpi = (\pi_{1}, \pi_{2},\ldots, \pi_{R})^{\prime}$, $\vtheta_{r} = (\mu_{r}, \sigma_{r}^{2})^{\prime}$, and $\mTheta = \left\{ \vtheta_{r} \right\}_{r = 1}^{R}$, where $\displaystyle \sum_{r = 1}^{R} \pi_{r} = 1$.
Then, the PDF of the MLN distribution with $R$ components is expressed by
\begin{equation}
	f(x | \vpi, \mTheta) = \sum_{r = 1}^{R} \pi_{r} f\left( x | \vtheta_{r} \right)= \sum_{r = 1}^{R} \frac{\pi_{r}}{\sqrt{2 \pi \sigma_{r}^{2}}x}\exp\left\{ -\frac{(\ln x - \mu_{r})^{2}}{2\sigma_{r}^{2}} \right\},
	\label{eq:PDF}
\end{equation}
and the CDF is expressed by
\begin{equation}
	F(x| \vpi, \mTheta) = \sum_{r = 1}^{R} \pi_{r} F(x | \vtheta_{r}) = \sum_{r = 1}^{R} \pi_{r} \Phi\left( \frac{\ln x - \mu_{r}}{\sigma_{r}} \right).
	\label{eq:CDF}
\end{equation}
If we substitute \eqref{eq:PDF} and \eqref{eq:CDF} for \eqref{eq:L}, it becomes the likelihood function for the MLN distribution model.
However, its Gini coefficient does not have a closed form.
Therefore, it is calculated from \eqref{eqn:gini}.
In the next section, we will consider the MLN distribution model with an unknown number of components, where $R$ is also treated as one of the parameters.

\section{Posterior Analysis\label{sec:PA}}
\subsection{Joint Posterior Distribution}
The likelihood function given in \eqref{eq:L} for the MLN distribution model is not particularly useful for Bayesian inference because its full conditional distributions are not the standard forms.
In this study, we consider an alternative approach based on the framework by \citet{GdL14}.
They proposed augmenting the model with vectors of latent variables $\vx = (x_{1}, x_{2}, \ldots, x_{n})^{\prime}$ and $x_{n_{k}^{*}}$ is set to $t_{k}$ for $k = 1,\ldots, K-1$, where $\displaystyle n_{k}^{*} = \sum_{j = 1}^{k}n_{j}$ and $\vz = (z_{1}, z_{2}, \ldots, z_{n})^{\prime}$, where $z_{i} = r \in \{1,2, \ldots, R\}$.
To complete this augmented likelihood, we also introduce a vector of observed variable $\displaystyle \vd = (\underbrace{1,1,\ldots,1}_{n_{1}},\ldots, \underbrace{k,k,\ldots,k}_{n_{k}},\ldots,\underbrace{K,K,\ldots,K}_{n_{K}})^{\prime}$, instead of $\vn$.
Then, the joint likelihood of $(\vx, \vd, \vz, \vpi, \mTheta)$ can be specified as
\begin{eqnarray}
	L(\vx, \vd, \vz, \vpi, \mTheta) \propto \prod_{r=1}^{R}\left( \sigma_{r}^{2} \right)^{-\frac{n_{r}}{2}} \pi^{n_{r}} \exp\left\{ -\sum_{i=1}^{n}\sum_{k=1}^{K}\frac{(\ln x_{i} - \mu_{r})^{2}}{2\sigma_{r}^{2}}I(d_{i} = k)I(z_{i} = r) \right\},
	\label{eq:AL}
\end{eqnarray}
where $n_{r} = \#\left\{ i: z_{i} = r \right\}$ and $I(A)$ denotes the indicator function of the event $A$.

As we adopt a Bayesian approach and extend the model to allow the number of components to change, we complete the model by specifying the following hierarchical prior distributions over the parameters ($R, \vpi, \mTheta$):
\begin{eqnarray*}
	R \sim \PO(\lambda_{0})I(R \le R_{\max}),\quad
	\vpi^{\prime} \sim \D(\underbrace{\alpha_{0}, \alpha_{0}, \ldots, \alpha_{0}}_{R}),\quad
	\mu_{r} | \mu, \tau^{2} \sim \N(\mu, \tau^{2}),\\
	\mu \sim \N(\mu_{0}, \tau_{0}^{2}),\quad
	\tau^{-2} \sim \G(n_{0}, s_{0}),\quad
	\sigma_{r}^{-2}| \beta \sim \G(\nu_{0}, \beta),\quad
	\beta \sim \G(g_{0}, h_{0}),
\end{eqnarray*}
where $\PO(\lambda)$ means a Poisson distribution and $I(R \le R_{\max})$ imposes the preassigned upper limit $R_{\max}$ on the number of components.
$\D(\alpha, \alpha, \ldots, \alpha)$ means a symmetric Dirichlet distribution and $\G(a,b)$ is a gamma distribution with scale and shape parameters $a$ and $b$, respectively.

It should be mentioned that the use of midpoint and range of data is recommended as the hyper-parameters in \citet{RG97}.
On the other hand, \citet{GdL14} used an improper prior.
For the grouped data in income distribution, it is difficult to find a midpoint and range of data.
Therefore, the hierarchical prior is assigned and they are also treated as parameters in the model to avoid it.

\subsection{Posterior Simulation}
As the joint posterior distribution is much simplified, we can now use MCMC methods.
The Markov chain sampling scheme can be constructed from birth-and-death process, split-or-combine process, and the full conditional distributions of $R, \vpi, \left\{ \mu_{r} \right\}, \left\{ \sigma_{r}^{2} \right\}, \left\{ z_{i} \right\}, \left\{ x_{i} \right\}_{i:i \ne n_{k}^{*}}, \mu, \tau^{2}, \beta$.

\subsubsection{Birth and Death Process}
To implement a birth and death process, we first make a random choice between birth and death with the probability $b_{R}$ and $d_{R}$, where $d_{R} = 1 - b_{R} = 0.5$ except for $d_{1} = 0$ and $b_{R_{\max}} = 0$.
For a birth process, a weight and parameters for the proposed new component are sampled from
\begin{eqnarray}
	\pi_{r^{*}} \sim \B(1, R),\quad \mu_{r^{*}} \sim \N(\mu, \tau^{2}),\quad \sigma_{r^{*}}^{-2} \sim \G(\nu_{0}, \beta),
\end{eqnarray}
where $\B(p,q)$ is a beta distribution.
For a death process, a random choice is made between any empty components and the chosen component is deleted.
Then, the acceptance probabilities $\min(1, A)$ and $\min(1, A^{-1})$ for birth and death are evaluated by
\begin{eqnarray}
	A = \frac{\pi(R+1)}{\pi(R)}\frac{1}{B(R\alpha_{0},\alpha_{0})}\pi_{r^{*}}^{\alpha_{0}-1}(1 - \pi_{r^{*}})^{n + R \alpha_{0} - R} (R + 1)\frac{d_{R+1}}{(R_{0} + 1) b_{R}}\frac{1}{g_{1,R}(\pi_{r^{*}})}(1-\pi_{r^{*}})^{R-1},
\end{eqnarray}
where $g_{p,q}$ denotes the $\B(p,q)$ density, $B(p,q)$ is a beta function and $R_{0}$ is the number of empty components \citep[see also][]{RG98}.

\subsubsection{Split or Combine Process}
Using the same probabilities $b_{R}$ and $d_{R}$ as above, we make a random choice between attempting to split or combine, depending on $R$.
Our combine proposal begins by choosing a pair of adjacent components ($r_{1}, r_{2}$) at random, which satisfies that there is no other $\mu_{r}$ in the interval $[\mu_{r_{1}}, \mu_{r_{2}}]$.
Then, the combined component, here labeled $r^{*}$, is created as ($\pi_{r^{*}}, \mu_{r^{*}}, \sigma_{r^{*}}^{2}$), which satisfies the following equations:
\begin{eqnarray}
	\pi_{r^{*}} &=& \pi_{r_{1}} + \pi_{r_{2}},\\
	\pi_{r^{*}} \mu_{r^{*}} &=& \pi_{r_{1}} \mu_{r_{1}} + \pi_{r_{2}} \mu_{r_{2}},\\
	\pi_{r^{*}} \left( \mu_{r^{*}}^{2} + \sigma_{r^{*}}^{2} \right) &=& \pi_{r_{1}} \left( \mu_{r_{1}}^{2} + \sigma_{r_{1}}^{2} \right) + \pi_{r_{2}} \left( \mu_{r_{2}}^{2} + \sigma_{r_{2}}^{2} \right).
\end{eqnarray}

To make a split proposal, we begin with choosing a component, here labeled $r^{*}$, and drawing a three-dimensional random variables as follows:
\begin{eqnarray}
	u_{1} \sim \B(2,2),\quad u_{2} \sim \B(2,2),\quad u_{3} \sim \B(1,1).
\end{eqnarray}
Then, the split proposal is made as follows:
\begin{eqnarray*}
	\pi_{r_{1}} = \pi_{r^{*}} u_{1}, && \pi_{r_{1}} = \pi_{r^{*}} (1 - u_{1}),\\
	\mu_{r_{1}} = \mu_{r^{*}} - u_{2} \sigma_{r^{*}} \sqrt{\frac{\pi_{r_{2}}}{\pi_{r_{1}}}}, && \mu_{r_{2}} = \mu_{r^{*}} + u_{2} \sigma_{r^{*}} \sqrt{\frac{\pi_{r_{1}}}{\pi_{r_{2}}}},\\
	\sigma_{r_{1}}^{2} = u_{3}(1-u_{2}^{2})\sigma_{r^{*}}^{2} \frac{\pi_{r^{*}}}{\pi_{r_{1}}}, && \sigma_{r_{2}}^{2} = (1 - u_{3})(1-u_{2}^{2})\sigma_{r^{*}}^{2} \frac{\pi_{r^{*}}}{\pi_{r_{2}}},
\end{eqnarray*}
where the adjacency condition that there is no other $\mu_{r}$ in the interval $[\mu_{r_{1}}, \mu_{r_{2}}]$ is satisfied.

Finally, the acceptance probability $\min(1, A)$ for split is evaluated  by
\begin{eqnarray}
	A &=& (\text{likelihood ratio})\frac{\pi(R+1)}{\pi(R)}(R + 1)\frac{\pi_{r_{1}}^{\alpha_{0} - 1 + l_{1}}\pi_{r_{2}}^{\alpha_{0} - 1 + l_{2}}}{\pi_{r^{*}}^{\alpha_{0} - 1 + l_{1} + l_{2}}B(\alpha_{0},R\alpha_{0})}\nonumber\\
	&& \times \frac{1}{\sqrt{2 \pi \tau^{2}}}\exp\left\{ -\frac{(\mu_{r_{1}} - \mu)^{2} + (\mu_{r_{2}} - \mu)^{2} - (\mu_{r^{*}} - \mu)^{2}}{2\tau^{2}} \right\}\nonumber\\
	&& \times \frac{\beta^{\nu_{0}}}{\Gamma(\nu_{0})}\left( \frac{\sigma_{r_{1}}^{2}\sigma_{r_{2}}^{2}}{\sigma_{r^{*}}^{2}} \right)^{-n_{0}-1}\exp\left\{ -\beta\left( \sigma_{r_{1}}^{-2} + \sigma_{r_{2}}^{-2} -\sigma_{r^{*}}^{-2} \right) \right\}\nonumber\\
	&& \times \frac{d_{R + 1}}{b_{R}P_{\text{alloc}}} \frac{1}{g_{2,2}(u_{1})g_{2,2}(u_{2})g_{1,1}(u_{3})}\nonumber\\
	&&\times \frac{\pi_{r^{*}}|\mu_{r_{1}} - \mu_{r_{2}}|\sigma_{r_{1}}^{2}\sigma_{r_{2}}^{2}}{u_{2}(1-u_{2})u_{3}(1-u_{3})\sigma_{r^{*}}^{2}},
\end{eqnarray}
where $l_{1}$ and $l_{2}$ are the numbers of observations proposed to be assigned to $r_{1}$ and $r_{2}$ and $P_{\text{alloc}}$ is the probability that this particular allocation is made.
For the corresponding combine move, the acceptance probability is $\min(1, A^{-1})$, using the same expression for $A$, but some obvious differences in the substitutions.

Unsurprisingly, we can use the same acceptance probabilities with \citet{RG97} in spite of the fact that this model includes additional parameters ($\mu,\ \tau^{2}$) and a latent vector ($\vx$).
This is because the additional parameters and a latent vector are independent from the number of components.
These two processes are explained in more detail in \citet{RG97}.

\subsubsection{Sampling the Other Parameters}
With the exception of some hyper-parameters, the other parameters are easily sampled from the standard distributions following \citet{DR94} and \citet{GdL14}.
The full conditional distribution for $\vpi$ remains Dirichlet in form:
\begin{eqnarray}
	\vpi^{\prime} | \cdots \sim \D(n_{1} + \alpha_{0}, \ldots, n_{R} + \alpha_{0}),
	\label{fcd:pi}
\end{eqnarray}
where we use `$|\cdots$' to denote conditioning on all other variables.

The full conditionals for $\left\{ \mu_{r} \right\}$ and $\left\{ \sigma_{r}^{2} \right\}$ are
\begin{eqnarray}
	\mu_{r} | \cdots \sim \N(\hat{\mu}_{r}, \hat{\tau}_{r}^{2}),\quad \sigma_{r}^{-2} | \cdots \sim \G(\hat{\nu}_{r}, \hat{\beta}_{r}),
	\label{fcd:mu-sigma2}
\end{eqnarray}
where $\displaystyle \hat{\tau}_{r}^{2} = \left( \sigma_{r}^{-2} n_{r} + \tau^{-2} \right)^{-1}$,
$\displaystyle \hat{\mu}_{r} = \hat{\tau}_{r}^{2} \left( \sigma_{r}^{-2} \sum_{i: z_{i} = r} \ln x_{i} + \tau^{-2} \mu \right)$,
$\hat{\nu}_{r} = 0.5 n_{r} + \nu_{0}$ and 
$\displaystyle \hat{\beta}_{r} = 0.5 \sum_{i: z_{i} = r}\left( \ln x_{i} - \mu_{r} \right)^{2} + \beta$.
Although \citet{LN16} considered several restrictions to avoid the label switching problem, we simply assume that $\mu_{1}<\mu_{2}\ldots<\mu_{R}$ to help remove the label switching problem \citep[see][]{DR94}.

For the allocation variables, we have
\footnote{
	It should be mentioned that \citet{GdL14} derived the full conditional distribution of (B) $z_{i} | \vpi, \mTheta, d_{i}$, which is one without the condition on $x_{i}$.
	However, it is not required to decide the initial values of $z_{i}$ and we can choose any $z_{i}$.
	The simplest example is to start from $R=1$.
	Even if we start from any $R > 1$, (B) in Step 2 in \citet{GdL14} is not required.
	We can start from any random $z_{i}$.
}
\begin{eqnarray}
	\pi(z_{i} = r|\cdots) \propto \frac{\pi_{r}}{\sigma_{r}} \exp\left\{ -\frac{(\ln x_{i} - \mu_{r})^{2}}{2 \sigma_{r}^{2}} \right\}.
	\label{fcd:z}
\end{eqnarray}

For the latent variables $x_{i}$, $i=1,2,\ldots,n$ except for $i = n_{k}^{*}$, $k = 1,2,\ldots, K-1$, the full conditional distributions are
\begin{eqnarray}
	x_{i} | \cdots \sim \LN(\mu_{r}, \sigma_{r}^{2})I(t_{d_{i}-1} < x_{i} \le t_{d_{i}}).
	\label{fcd:x}
\end{eqnarray}

For the hyper-parameters that we are not treating as fixed, $\mu$, $\tau^{2}$ and $\beta$, have
\begin{eqnarray}
	\mu | \cdots \sim \N(\hat{\mu}, \hat{\tau}^{2}),\quad \tau^{-2}|\cdots\sim \G(\hat{n},\hat{s}),\quad \beta | \cdots \sim \G(\hat{g}, \hat{h} ),
\end{eqnarray}
where
$\hat{\tau}^{2} = (\tau^{-2}R + \tau_{0}^{-2})^{-1}$,
$\displaystyle \hat{\mu} = \hat{\tau}^{2}(\tau^{-2}\sum_{r=1}^{R} \mu_{r} + \tau_{0}^{-2} \mu_{0})$,
$\hat{g} = R \nu_{0} + g_{0}$,
$\displaystyle \hat{h} = \sum_{r=1}^{R}\sigma_{r}^{-2} + h_{0}$.

\section{Numerical Examples by Simulated Data}\label{sec:NESD}
To illustrate the Bayesian approach discussed in the previous section, we consider two simulated data examples.
One is the case where the true data generating process (DGP) is the MLN distribution, and the other is the case where the true DGP is the generalized beta distribution of the second kind (GB2 distribution).
In the first example, we examine the performance of our method and compare the GB2 distribution.
In the second example, we explore the possibility of the MLN distribution model assuming that the true DGP is the GB2 distribution.
The reason for choosing the GB2 distribution as the competing distribution against the MLN distribution is that the GB2 distribution is reported to fit the data well in the empirical analyses.
Therefore, it is worthwhile to examine the fit of the MLN distribution when the true DGP is the GB2 distribution.
All the results reported here were generated using Ox version 9.30 (macOS\_64/Parallel) \citep[see][]{D13} and all the figures are drawn using R version 4.5.1 \citep[see][]{R25}.

\subsection{Example 1}

\begin{figure}[tbp]
	\centering
	\includegraphics[width=.5\linewidth]{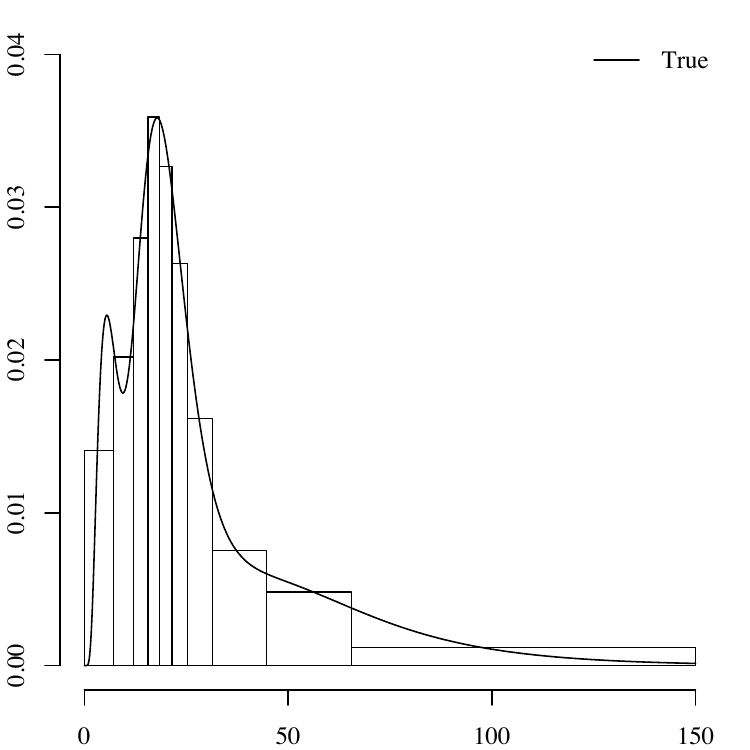}
	\caption{Simulated data 1: The histogram and true distribution}
	\label{fig:sim1-hist}
\end{figure}

\begin{center}[INCLUDE \autoref{fig:sim1-hist} HERE]\end{center}

In the first simulated example, we examine the performance of our algorithm, and then compare the distribution with the GB2 distribution.
The simulated data, wherein the DGP is the MLN distribution, is generated as follows.
First, $x_{i}$, $i = 1,\ldots, 10,000$ were generated from the MLN distribution with three components ($R =3$) with parameters $\vpi = (0.2,\ 0.5,\ 0.3)^{\prime}$, $\vmu = (2.0,\ 3.0,\ 4.0)^{\prime}$ and $\vsigma^{2} = (0.3,\ 0.1,\ 0.2)^{\prime}$.
The generated random numbers are sorted in ascending order, and $x_{n_{k}}$ corresponds to the $n_{k}$th observation, where $\displaystyle n_{k} = n \times \frac{k}{K}$.
Then, $k=1,2,\ldots,K-1$ is picked up and $\vt = (t_{1}, t_{2},\ldots,t_{K-1})^{\prime}$, where $t_{k} = x_{n_{k}}$, are collected.
In this example, $K$ is set to $10$, which means that the dataset is decile data.
Figure \ref{fig:sim1-hist} shows the true distribution and histogram, which is drawn from the simulated data.
From the figure, we can observe the following features under this setting: 
(i) the first and second components' modes are found easily whereas the third one is not in the true distribution;
(ii) the histogram looks like a unimodal distribution with heavy tail, that is, it is difficult to identify the first component as well from the grouped data.

To proceed with the Bayesian analysis, we need to set the hyper-parameters.
For the prior distributions, we set the hyper-parameters as follows:
\begin{eqnarray*}
	\lambda_{0} = 10,\quad R_{\max} = 50,\quad \alpha_{0} = 1.0 ,\quad \mu_{0} = 0.0,\quad \tau_{0}^{2} = 100.0,\\
	n_{0} = 2.0,\quad s_{0} = 0.01,\quad \nu_{0} = 2.0,\quad g_{0} = 0.2,\quad h_{0} = 0.01.
\end{eqnarray*}
With the simulated data, we ran the MCMC algorithm using $500,000$ and discarding the first $100,000$ iterations.

\begin{figure}[tbp]
	\centering
	\includegraphics[width=.5\linewidth]{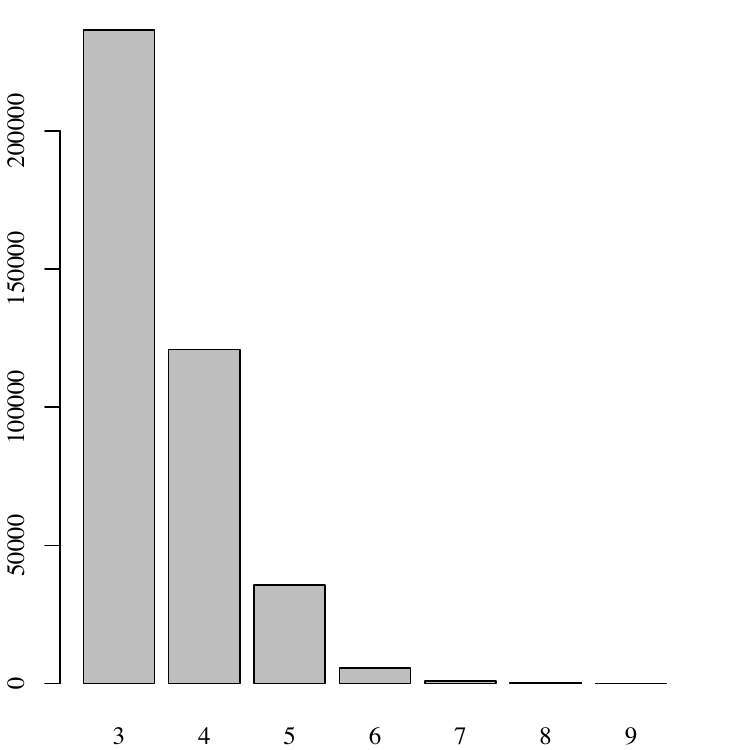}
\caption{Simulated data 1: Posterior distribution of $R$}
	\label{fig:sim1-r}
\end{figure}

\begin{center}[INCLUDE \autoref{fig:sim1-r} HERE]\end{center}

\begin{figure}[tbp]
	\centering
	\includegraphics[width=.5\linewidth]{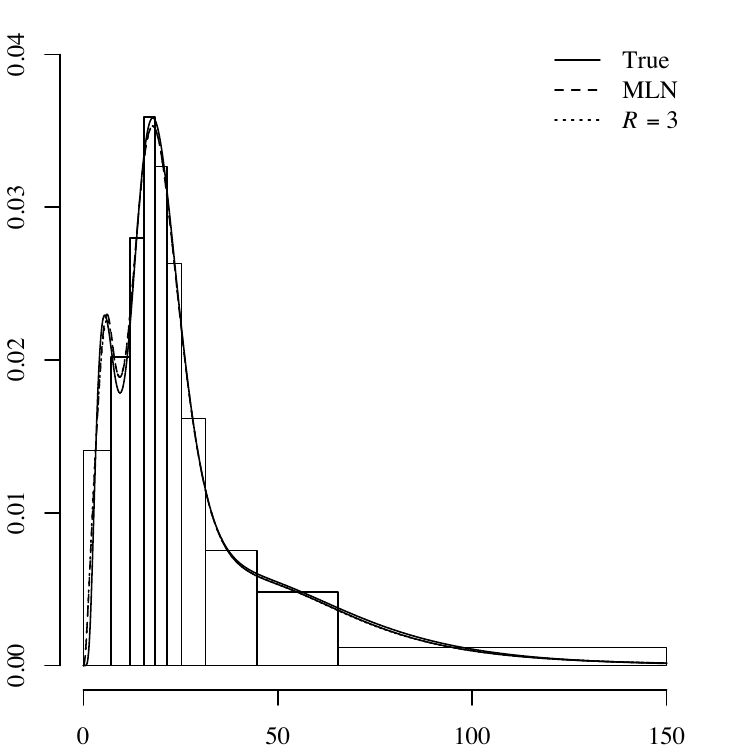}
	\begin{minipage}{.8\textwidth}
		Note:
		The histogram, true distribution, unconditional predictive distribution for the MLN distribution (MLN), and conditional predictive distribution for the MLN distribution on $R=3$ ($R=3$) are drawn. 
	\end{minipage}
	\caption{Simulated data 1: The histogram, true distribution, and predictive distributions}
	\label{fig:sim1-density1}
\end{figure}

\begin{center}[INCLUDE \autoref{fig:sim1-density1} HERE]\end{center}

\begin{center}[INCLUDE \autoref{tab:sim} HERE]\end{center}

Figure \ref{fig:sim1-r} shows the posterior distribution of $R$.
From the figure, we can confirm that the true number of components $R = 3$ shows the highest posterior mass.
Therefore, we will focus on the result of the conditional posterior results on $R=3$ hereafter, if we report the result of a conditional one. We can also conclude that our algorithm can identify the exact number of components.
Figure \ref{fig:sim1-density1} shows the unconditional predictive distribution and the predictive distribution conditioned on $R=3$.
From the figure, we find that the unconditional predictive distribution and the predictive distribution conditioned on $R=3$ show the similar shape.
In addition, they also show a similar shape with the true distribution, although there is a slight difference around the first component, where the first mode of the predictive distribution is lower than that of the true distribution.
To see the difference between the predictive and true distributions, Table \ref{tab:sim} shows the conditional posterior estimates on $R=3$.
From the table, we can see that all the estimates include the true values in the 95\% credible intervals and most of the posterior means are close to the true values. However, it seems difficult to identify the first component because the standard deviations of the first component are larger than those of other components, and the 95\% credible intervals of the first component are wider than those of others.
Nevertheless, we can conclude that our method not only identifies the true number of components, but also estimates the parameters accurately.

\begin{center}[INCLUDE \autoref{tab:sim1-ML} HERE]\end{center}

\begin{figure}[tbp]
	\centering
	\includegraphics[width=.5\linewidth]{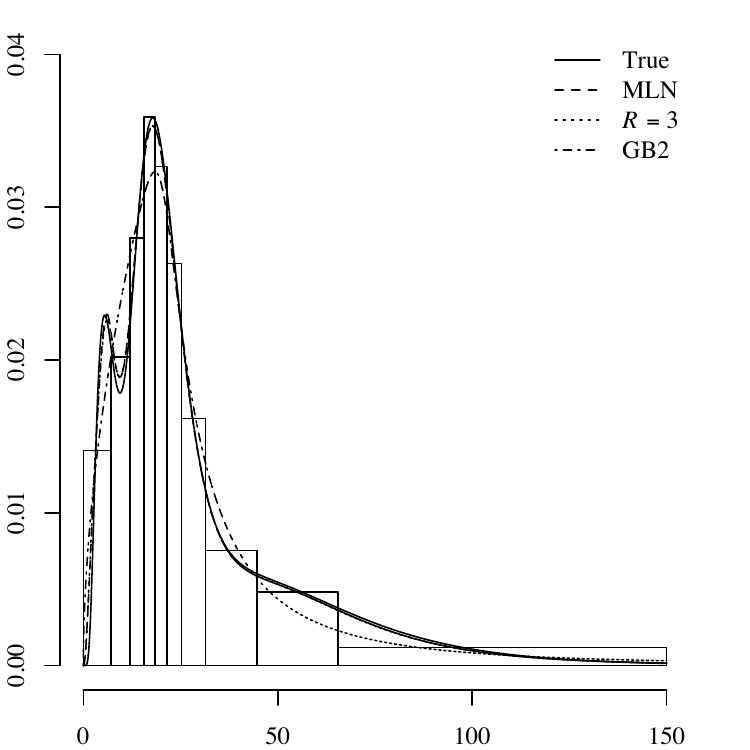}
	\begin{minipage}{.8\textwidth}
		Note:
		The histogram, true distribution, unconditional predictive distribution for the MLN distribution (MLN), conditional predictive distribution for the MLN distribution on $R=3$ ($R=3$), and predictive distribution for the GB2 distribution (GB2) are drawn. 
	\end{minipage}
	\caption{Simulated data 1: The histogram, true distribution, and predictive distributions}
	\label{fig:sim1-density2}
\end{figure}

\begin{center}[INCLUDE \autoref{fig:sim1-density2} HERE]\end{center}

Based on the favorable performance of the model and our algorithm in the simulated data, we can consider the goodness-of-fit of this model.
\citet{BMM97,KN19} confirmed that the fit of the GB2 distribution proposed by \citet{M84} performed quite well in the empirical analyses.
Therefore, we examine the goodness-of-fit of the GB2 distribution model when the true distribution is the MLN distribution model.
The parameters of the GB2 distribution is estimated using the Tailored randomized block Metropolis-Hastings (TaRBMH) algorithm by \citet{KN19}, which is first proposed by \citet{CR10} for estimating the DSGE model, and we examine the goodness-of-fit using the marginal likelihoods in line with \citet{KN19}.
Table \ref{tab:sim1-ML} shows the log of marginal likelihoods for these distributions, which are calculated by the harmonic mean estimator proposed by \citet{NR94} for its simplicity.
\footnote{
	The parameters and the harmonic mean estimate from GB2 distribution are estimated independently from the MLN distribution and we ran the MCMC algorithm using $40,000$ and discarding the first $10,000$ iterations.
	For the parameters $a$, $b$, $p$, $q$ in the GB2 distribution, we assume the following prior distributions:
	\begin{align*}
		a \sim \G(\alpha_{0},\beta_{0}),\quad b \sim \G(\gamma_{0},\delta_{0}),\quad p \sim \G(\epsilon_{0},\zeta_{0}),\quad q \sim \G(\eta_{0},\theta_{0}),
	\end{align*}
	and the hyper-parameters are set to $\alpha_{0} = \beta_{0} = \gamma_{0} = \delta_{0} = \epsilon_{0} = \zeta_{0} = \eta_{0} = \theta_{0} = 1.0$.
}
From the table, we can confirm that the unconditional MLN distribution model shows the highest and the GB2 distribution model shows the lowest log of marginal likelihood.
The log of marginal likelihood of the conditional MLN distribution on $R=3$ lies between the unconditional MLN and GB2 distribution models and is much higher than that of GB2 distribution.
Figure \ref{fig:sim1-density2} shows the unconditional predictive distributions for the MLN distribution, conditional predictive distribution for the MLN distribution on $R=3$, and predictive distribution for the GB2 distributions.
From the figure, we can confirm that the unconditional predictive distribution for the MLN distribution and the conditional predictive distribution for the MLN distribution on $R=3$ seem to be similar to the true distribution, whereas the predictive distribution for the GB2 distribution seems to be different from the true distribution.
Therefore, we can conclude that the unconditional predictive distribution for the MLN distribution model is a good fit for the income distribution, and, that the conditional predictive distribution, which includes the number of components, is useful for interpretation in an economically meaningful way.

\begin{figure}[tbp]
	\centering
	\includegraphics[width=.5\linewidth]{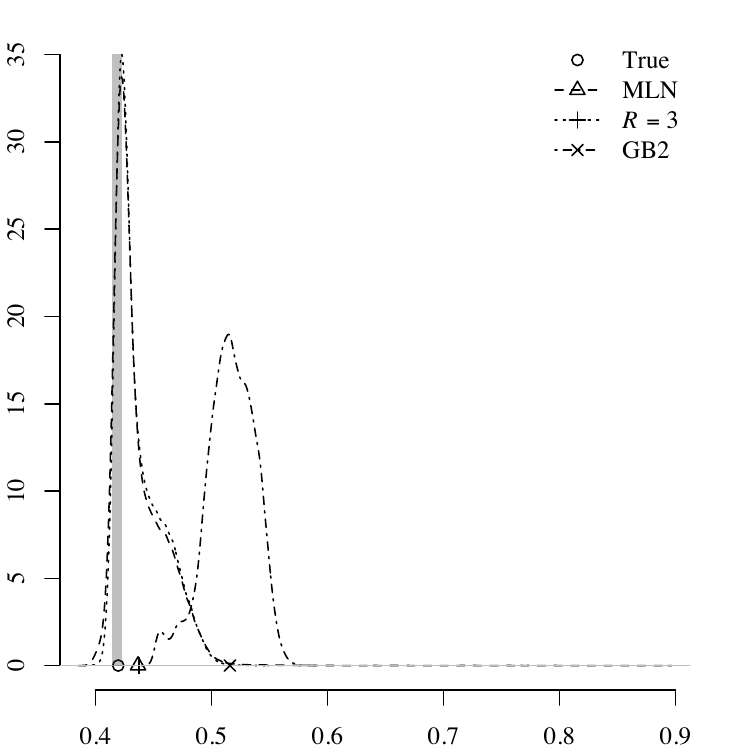}
	\caption{Simulated data 1: The posterior distributions of the Gini coefficients}
	\label{fig:sim1-gini}
\end{figure}

\begin{center}[INCLUDE \autoref{fig:sim1-gini} HERE]\end{center}

\begin{center}[INCLUDE \autoref{tab:sim1-gini} HERE]\end{center}

Once the parameters are estimated, the Gini coefficient can be calculated from the parameters using \eqref{eqn:gini}.
To confirm the accuracy of the estimates, we calculated the Gini coefficient by means of the numerical integration.
Figure \ref{fig:sim1-gini} shows the posterior distributions of the Gini coefficients for the unconditional MLN distribution, the conditional MLN distribution on $R=3$, and the GB2 distribution.
The posterior summaries are also reported in Table \ref{tab:sim1-gini}.
To see the accuracy of the Gini coefficients, the nonparametric lower and upper bounds of the Gini coefficient ($0.4144, 0.4226$), which was proposed by \citet{G72}, are shaded in the figure.
From the results, we can confirm that not only all of the $95$\% credible intervals are wider than the nonparametric bound, but also the posterior means are estimated outside the bound.
Moreover, the posterior means for the unconditional MLN distribution model and the conditional MLN distribution on $R=3$ are similar, whereas the posterior mean for the GB2 distribution is farther than these means from the true value ($0.4196$).
Simultaneously, we can see that the posterior modes for both the unconditional and conditional MLN distribution models on $R=3$ seem to be close to the true value.
Therefore, we also calculated the posterior modes using the half sample mode estimator by \citet{RC74}.
From the results, we can confirm that the posterior modes for the unconditional MLN distribution are estimated accurately.
Therefore, we can conclude that our algorithm can accurately estimate not only the number of components, but also the posterior estimates including the Gini coefficient, which is constructed as the function of the parameters.

\subsection{Example 2}

\begin{figure}[tbp]
	\centering
	\includegraphics[width=.45\linewidth]{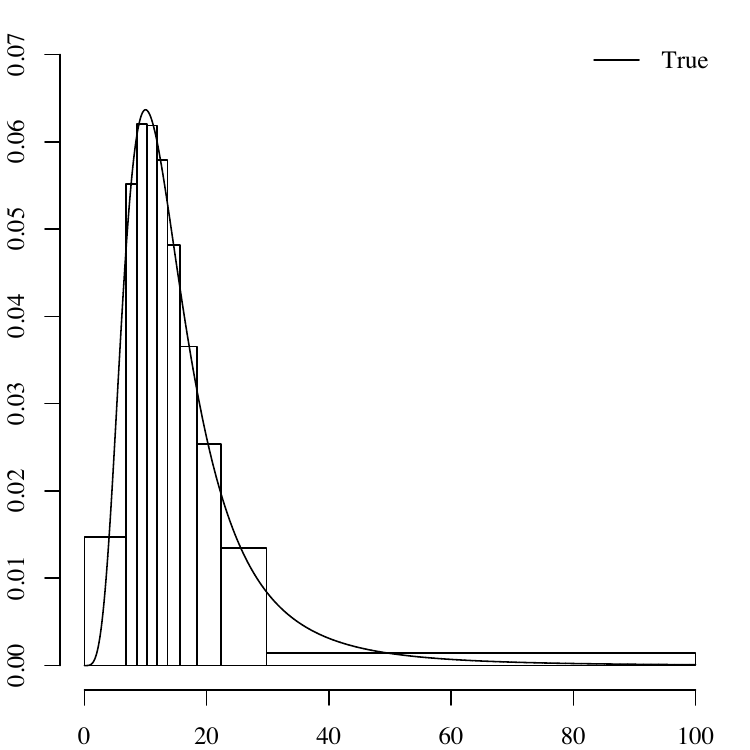}
	\caption{Simulated data 2: The histogram and true distribution}
	\label{fig:sim2-hist}
\end{figure}

\begin{center}[INCLUDE \autoref{fig:sim2-hist} HERE]\end{center}

In the previous subsection, we confirm that our algorithm can identify the true number of components and that the Gini coefficient can be calculated accurately.
In this subsection, we will consider the contrary situation, wherein the true distribution is different from the MLN distribution.
As we have found that when the MLN distribution is the true distribution, the GB2 distribution may not fit the distribution well or the calculation of the Gini coefficient may not be accurate, we explore the possibility of the MLN distribution model by considering the contrary case.
The simulated data, wherein the DGP is the GB2 distribution, is generated as follows.
First, $x_{i}$, $i = 1,\ldots, 10,000$ were generated from the GB2 distribution with parameters $a = 2.0$, $b=10.0$, $p=2.5$, and $q=1.5$.
The generated random numbers are sorted in ascending order, and $x_{n_{k}}$ corresponds to the $n_{k}$th observation, where $\displaystyle n_{k} = n \times \frac{k}{K}$.
Then, $k=1,2,\ldots,K-1$ is picked up and $\vt = (t_{1}, t_{2},\ldots,t_{K-1})^{\prime}$, where $t_{k} = x_{n_{k}}$, are collected.
In this example, $K$ is also set to $10$.
Figure \ref{fig:sim2-hist} shows the true distribution and histogram, which is drawn from the simulated data.
From the figure, we can confirm that the shape of the distribution exhibits the standard shape for income distribution, that is, it is unimodal and is right-skewed.
Using the same hyper-parameters with the previous subsection, we ran the MCMC algorithm using $500,000$ and discarding the first $100,000$ iterations.
\footnote{
	For the case of the GB2 distribution, we ran the MCMC algorithm using $50,000$ and discarding the first $10,000$ iterations.
}

\begin{figure}[tbp]
	\centering
	\includegraphics[width=.45\linewidth]{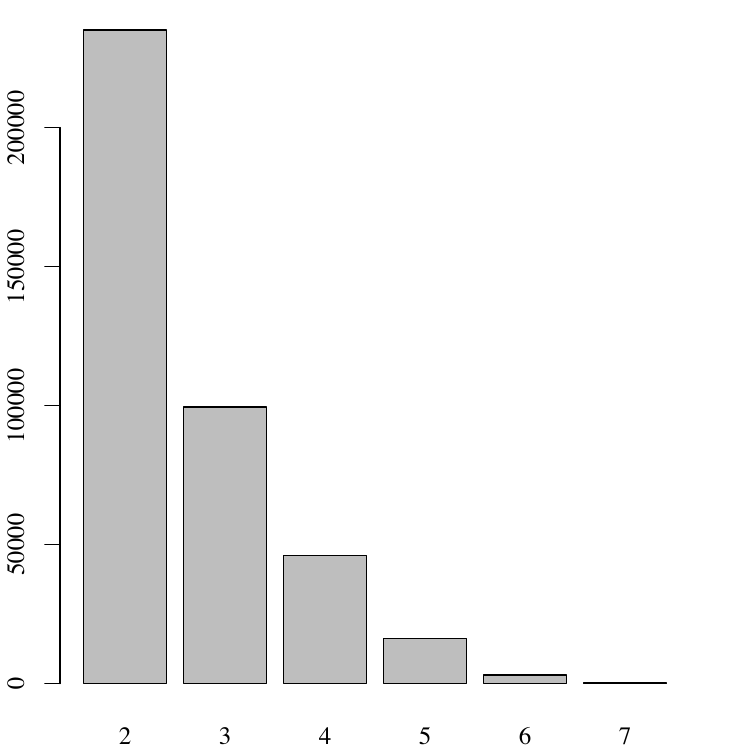}
\caption{Simulated data 2: Posterior distribution of $R$}
	\label{fig:sim2-r}
\end{figure}

\begin{center}[INCLUDE \autoref{fig:sim2-r} HERE]\end{center}

\begin{center}[INCLUDE \autoref{tab:sim2-ML} HERE]\end{center}

To keep this paper focused on the results of the MLN distribution model, we do not report the estimation result of the GB2 distribution; however, it should be mentioned that it was estimated quite well.
Figure \ref{fig:sim2-r} shows the posterior distribution of $R$ and the number of components $R=2$ shows the highest posterior mass.
Therefore, we will focus on the result of the conditional posterior result on $R=2$ hereafter.
As with the previous subsection, we will examine the goodness-of-fit of the distributions to examine the possibility of the MLN distribution model.
Table \ref{tab:sim2-ML} shows the log of marginal likelihoods for these distributions.
From the table, we can confirm that the log of marginal likelihood of the GB2 distribution model shows the highest value.
However, that of the unconditional MLN distribution is not so different if we take the standard error into account.
Conversely, that of the conditional MLN distribution model on $R=2$ indicates a smaller value than these distribution models.
Therefore, if we only focus on the fit of the distribution, the unconditional MLN distribution model becomes the alternative candidate against the GB2 distribution model.
However, if we are interested in the economically meaningful interpretation, the marginal likelihood can distinguish which distribution is preferred, because the log of marginal likelihood of the conditional MLN distribution model is smaller than that of the GB2 distribution.

\begin{figure}[tbp]
	\centering
	\includegraphics[width=.45\linewidth]{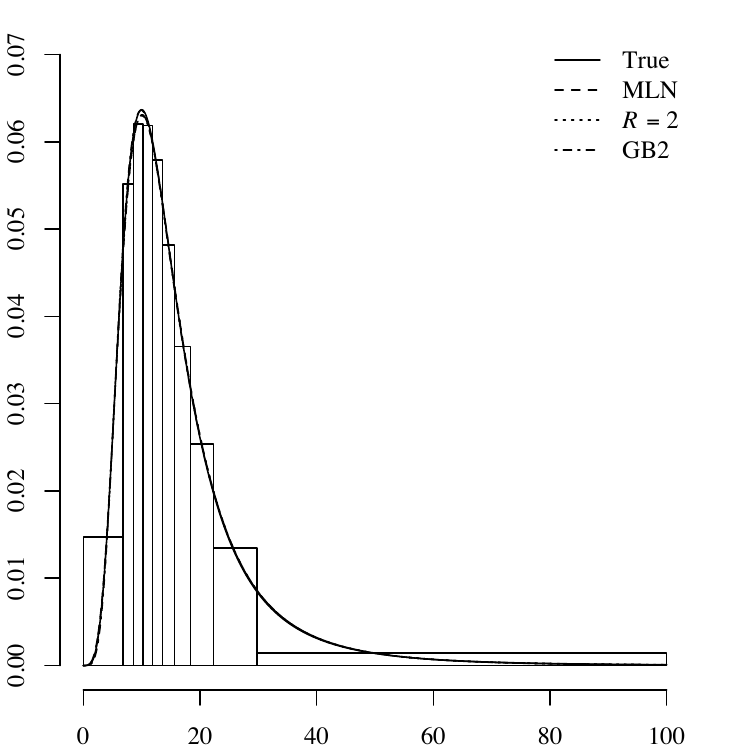}
	\begin{minipage}{.8\textwidth}
		Note:
		The histogram, true distribution, unconditional predictive distribution for the MLN distribution (MLN), conditional predictive distribution for the MLN distribution on $R=2$ ($R=2$), and predictive distribution for the GB2 distribution (GB2) are drawn. 
	\end{minipage}
	\caption{Simulated data 2: The histogram, true distribution, and predictive distributions}
	\label{fig:sim2-density2}
\end{figure}

\begin{center}[INCLUDE \autoref{fig:sim2-density2} HERE]\end{center}

Figure \ref{fig:sim2-density2} shows the posterior predictive distributions for these distribution models.
From the figure, we can see that all the predictive distributions are close to the true distribution.
However, we can also confirm that the conditional posterior predictive distribution on $R=2$ exhibits a small difference from the true distribution and that difference may lead to the difference in the log of marginal likelihoods.
Therefore, the choice of the distribution is very important and the marginal likelihoods play an important role to examine the goodness of fit.

\begin{figure}[tbp]
	\centering
	\includegraphics[width=.45\linewidth]{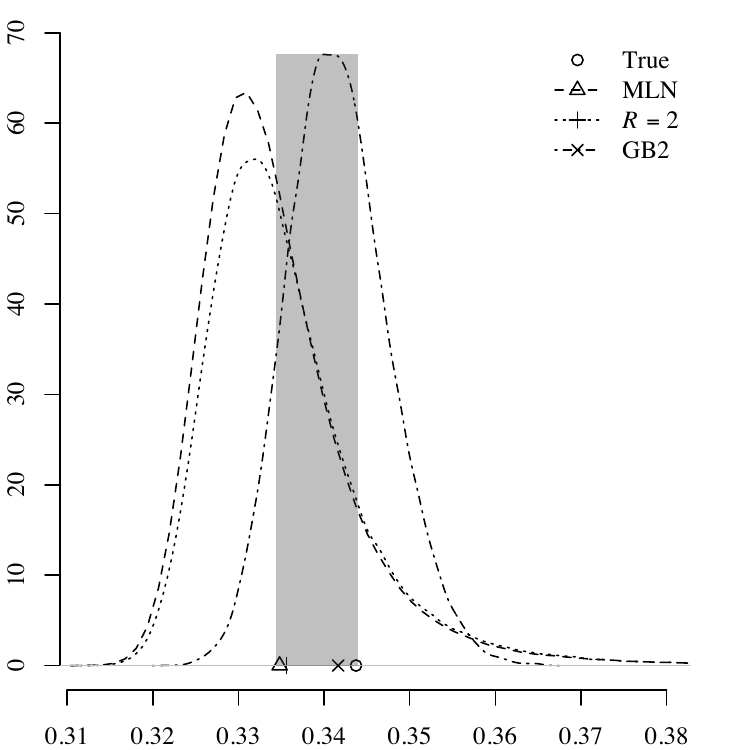}
	\caption{Simulated data 2: The posterior distributions of the Gini coefficients }
	\label{fig:sim2-gini}
\end{figure}

\begin{center}[INCLUDE \autoref{fig:sim2-gini} HERE]\end{center}

\begin{center}[INCLUDE \autoref{tab:sim2-gini} HERE]\end{center}

Finally, we examine the accuracy of the Gini coefficients.
Figure \ref{fig:sim2-gini} shows the posterior distributions of the Gini coefficients for the unconditional MLN distribution, conditional MLN distribution on $R=2$, and GB2 distribution.
The posterior summaries are also reported in Table \ref{tab:sim2-gini}.
To see the accuracy of the Gini coefficients, the nonparametric lower and upper bounds of the Gini coefficients ($0.3343, 0.3439$) are shaded in the figure.
From the results, we can confirm that all of the posterior means are included in the nonparametric bound, although all of the $95$\% credible intervals are wider than the nonparametric bound.
Moreover, the posterior mean for the GB2 distribution is closest to the true value ($0.3438$).
However, it should be mentioned that posterior modes for the unconditional and conditional MLN distribution models are estimated outside the bound.
Therefore, we can conclude that the Gini coefficients from both distributions can be calculated accurately.
However, those form these distributions infer to the GB2 distribution in terms of accuracy if the true DGP is the GB2 distribution.

\section{Applications to Real Data\label{sec:ARD}}
Using the Japanese household survey, Family Income and Expenditure Survey (FIES) in 2020, which is compiled by the Statistics Bureau of the Ministry of Internal Affairs and Communications, we will consider the income distributions and income inequalities in Japan.
The survey offers data, presented in Table 3, for two types of households: Yearly Average of Monthly Receipts and Disbursements per Household by Yearly Income Quintile Group, and by Yearly Income Decile Group (Two-or-more-person Households). 
Yearly pre-tax income is surveyed for two types of households, depending on the occupation of the households' head: workers' households are employed as clerks or wage earners by public or private enterprises, such as government office, private companies, factories, schools, hospitals, shops, etc, whereas two-or-more person households include those other than workers' households, such as individual proprietors households and households whose heads are merchants, artisans or administrators of unincorporated enterprise.
The sample size for each dataset is $n = 10,000$ and the dataset in decile form is utilized, therefore $n_{k} = 1,000$ for $k = 1,2,\ldots,K=10$.
Thus, only $\vt$ (unit: million yen) is different in each dataset.

\begin{figure}[tbp]
	\centering
	\includegraphics[width=.45\linewidth]{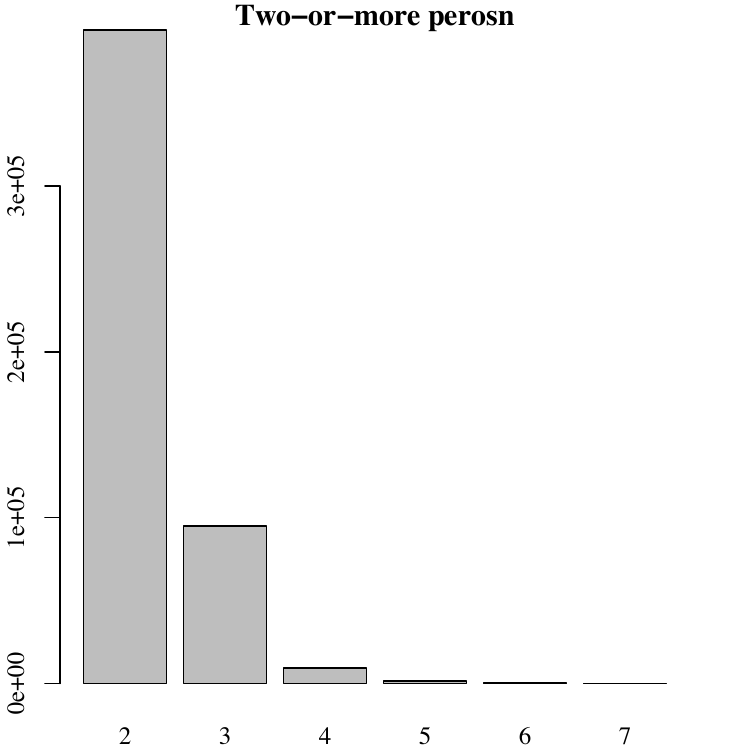}
	\includegraphics[width=.45\linewidth]{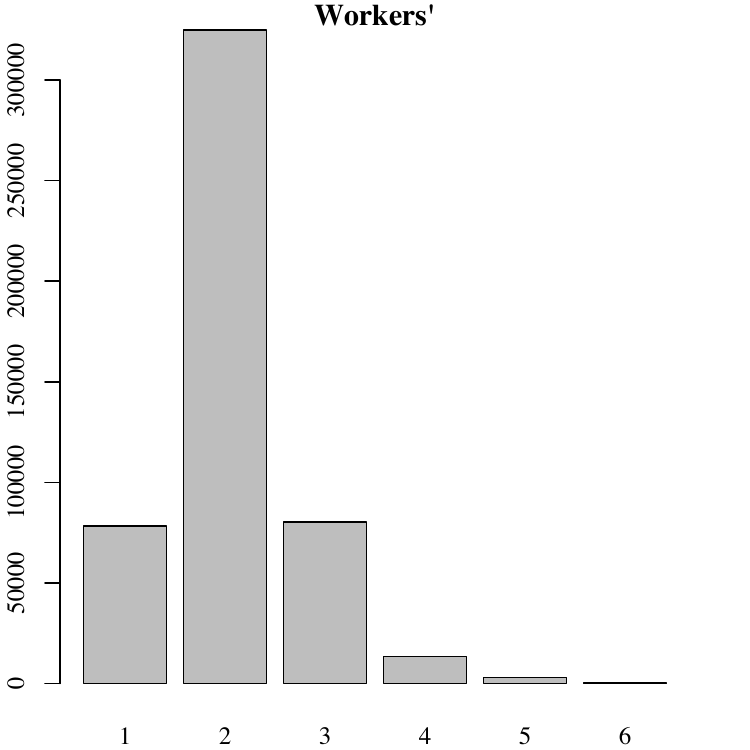}
	\caption{Posterior distributions of $R$}
	\label{fig:Emp-R}
\end{figure}

\begin{center}[INCLUDE \autoref{fig:Emp-R} HERE]\end{center}

\begin{center}[INCLUDE \autoref{tab:emp-ml} HERE]\end{center}

Using the same hyper-parameters as in the previous section, we ran the MCMC algorithm using $500,000$ iteration and discarding the first $100,000$ iterations for each dataset.
Figure \ref{fig:Emp-R} shows the posterior distribution of $R$ and we can see that the posteriors for $R$ favor the model with 2 components both for two-or-more person households and for workers' households.
Therefore, we proceed our discussion based on the results from $R=2$ both for two-or-more person households and for workers' households, when we examine the conditional ones.
Table \ref{tab:emp-ml} shows the log of marginal likelihoods for the unconditional MLN, conditional MLN on $R=2$, and GB2 distributions for both datasets.
From the table, we can confirm that the unconditional and conditional MLN distributions are preferred to the GB2 distribution.
Therefore, it allows us to interpret the parameters in an economically meaningful way using the results from the conditional MLN distribution on $R=2$.
The results from two-or-more person households and for workers' households suggest that there are two groups: a lower and higher income group.
In other words, Japanese households are divided into two groups both for two-or-more person households and for the workers' households, when the MLN distribution is assumed.
However, the interpretation of the parameters is different in each dataset.

\begin{center}[INCLUDE \autoref{tab:emp} HERE]\end{center}

Table \ref{tab:emp} shows the posterior estimates both for two-or-more person households and for workers' households.
From the table, we can make the following observations:
that 35.5\% households belong to the lower income group ($r=1$) in two-or-more person households from $\vpi$, whereas 64.8\% households belong to the lower income group in workers' households;
if we focus on the posterior estimates of $\vmu$, the posterior estimate $\mu_{2}$ of two-or-more person households is close to that of workers' households, whereas the posterior estimate $\mu_{1}$ of two-or-more person households is much smaller than that of workers' households;
and the posterior estimate $\sigma_{1}^{2}$ of two-or-more person households is smaller than $\sigma_{2}^{2}$, whereas $\sigma_{1}^{2}$ of workers' households is larger than $\sigma_{2}^{2}$.
As is described in the data explanation, workers' households are a subset of two-or-more person households.
Therefore, we can guess that the households, which are not included in the workers' households, make the meaning of the components different.
Even if the two components in workers' households are interpreted as lower and higher income class, the components wherein the households belong in workers' households might be different from the components wherein the households belong in two-or-more person households, because most of the estimates are different between two-or-more person and workers' households.
In addition to these interpretations, these differences may lead to the difference in the shape of income distribution.

\begin{figure}[tbp]
	\centering
	\includegraphics[width=.45\linewidth]{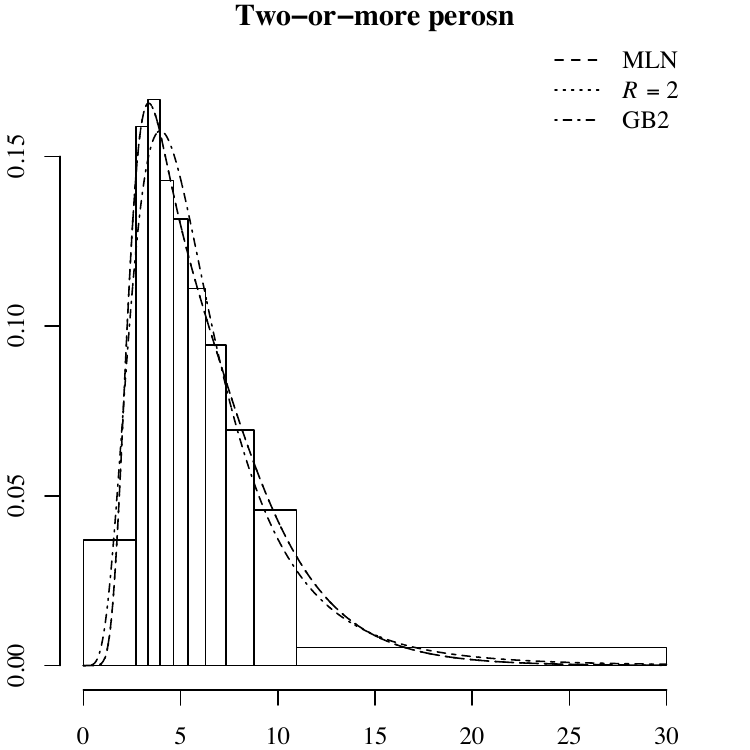}
	\includegraphics[width=.45\linewidth]{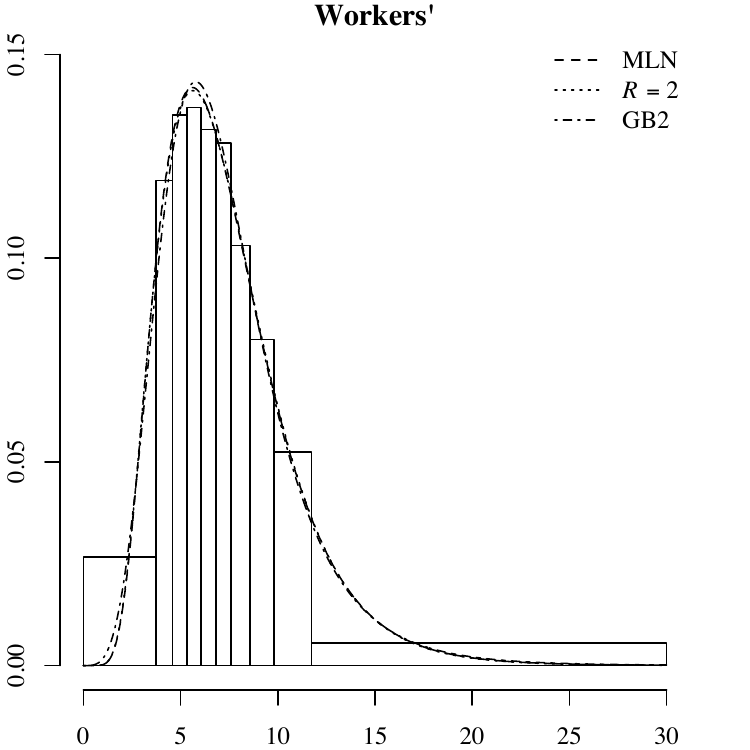}
	\begin{minipage}{.8\textwidth}
		Note:
		The histogram, unconditional predictive distribution for the MLN distribution (MLN), conditional predictive distribution for the MLN distribution on $R=2$ ($R=2$), and predictive distribution for the GB2 distribution (GB2) are drawn. 
	\end{minipage}
	\caption{The histogram and predictive distributions}
	\label{fig:Emp-Predictive}
\end{figure}

\begin{center}[INCLUDE \autoref{fig:Emp-Predictive} HERE]\end{center}

Figure \ref{fig:Emp-Predictive} shows the histogram from the data, predictive distributions for the unconditional and conditional MLN distributions, and GB2 distribution both for two-or-more person and workers' households.
First, we can confirm that the predictive distributions for the unconditional and conditional MLN distributions both from two-or-more person and workers' households are very similar. They seem to fit to the histograms, whereas those from the GB2 distribution are different from those from the unconditional and conditional MLN distributions, especially in the case of two-or-more person households.
Therefore, the goodness-of-fit of the MLN distribution model seems to be adequate for the Japanese income data.

\begin{center}[INCLUDE \autoref{tab:emp-gini} HERE]\end{center}

\begin{figure}[tbp]
	\centering
	\includegraphics[width=.45\linewidth]{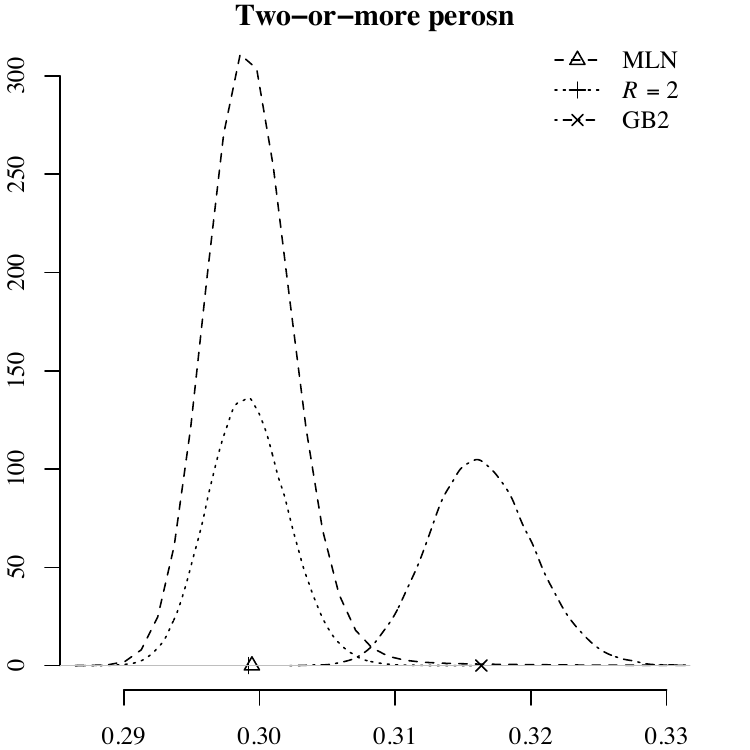}
	\includegraphics[width=.45\linewidth]{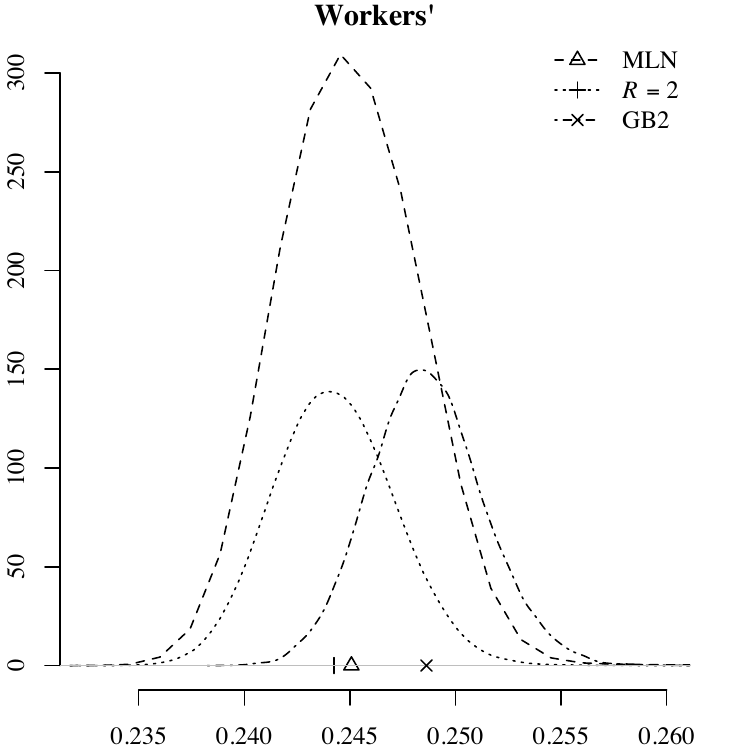}
	\caption{The posterior distributions of the Gini coefficients}
	\label{fig:Emp-Gini}
\end{figure}

\begin{center}[INCLUDE \autoref{fig:Emp-Gini} HERE]\end{center}

Finally, as the predictive distributions for both datasets seem to fit to the histograms, we are also estimate the Gini coefficients as is discussed in the previous section.
This is because the Gini coefficients are sometimes used for policy making and related matters.
To examine the features of the Gini coefficients, the estimated Gini coefficients are shown in Table \ref{tab:emp-gini} with the posterior distributions shown in Figure \ref{fig:Emp-Gini}.
At first, we can confirm that the Gini coefficient from two-or-more person households is larger than that of workers' households.
This might be caused by the additional households, which do not appear in the workers' households.
Secondly, the 95\% credible intervals do not overlap between the unconditional and conditional MLN distribution and GB2 distribution for two-or-more person households, whereas they overlap in the case of workers' households.
This may suggest that the choice of the distribution in two-or-more person households is more pronounced than that of workers' households. Furthermore, if we assume the GB2 distribution as the hypothetical income distribution, the Gini coefficients are overestimated both in two-or-more person and workers' households.
However, the log of marginal likelihood of the GB2 distribution is much smaller than that of the MLN distribution.
Therefore, we can avoid such an overestimation, if the marginal likelihoods are appropriately utilized.

\section{Conclusions\label{sec:C}}
There is a strong argument for employing a reversible jump MCMC algorithm for the MLN distribution model with an unknown number of components from grouped data.
Based on the simulated data examples, our proposed algorithm worked well in terms of fitting the distribution and enabled us to calculate the Gini coefficient accurately.
The unconditional MLN distribution model is useful if we are interested in the fit of the income distribution, whereas the conditional MLN distribution model is useful if we are interested in an economically meaningful interpretation.
A major strength of the reversible jump MCMC algorithm is that it can provide both results simultaneously in one estimation. This, along with the ability of the marginal likelihood to choose an appropriate distribution, makes it an algorithm of choice for estimating the MLN model.

The robust results support the case for using the MLN distribution model to compare other candidate distributions.
Finally, using FIES datasets in 2020, the income distributions and inequalities in Japan were examined.
The results indicated two subgroups, both in two-or-more person households and in workers' households.
However, the meanings of the two subgroups might be different in each dataset.
We also observed that the Gini coefficient of two-or-more person households are larger than that of workers' households.
Moreover, if we calculate the Gini coefficients from the GB2 distribution, the Gini coefficients are overestimated.

Finally, we discuss the remaining issue.
Although a reversible jump MCMC algorithm for grouped data is considered to determine the number of components, more sophisticated algorithms, which can determine the number of components, are proposed, for example, by \citet{MFG16}.
We need to examine more efficient algorithm, but our finding that a reversible jump MCMC algorithm can identify the number of components correctly even from grouped data, represents an interesting first step.

\bibliographystyle{jae}
\bibliography{references}

\begin{thebibliography}{34}
\expandafter\ifx\csname natexlab\endcsname\relax\def\natexlab#1{#1}\fi
\expandafter\ifx\csname url\endcsname\relax
  \def\url#1{\texttt{#1}}\fi
\expandafter\ifx\csname urlprefix\endcsname\relax\def\urlprefix{URL }\fi

\bibitem[{Bordley et~al.(1997)Bordley, McDonald and Mantrala}]{BMM97}
Bordley RF, McDonald JB, Mantrala A. 1997.
\newblock Something new, something old: Parametric models for the size of
  distribution of income.
\newblock \emph{Journal of Income Distribution} \textbf{6}: 91--103.

\bibitem[{Chib(1995)}]{C95}
Chib S. 1995.
\newblock Marginal likelihood from the {Gibbs} output.
\newblock \emph{Journal of the American Statistical Association} \textbf{90}:
  1313--1321.

\bibitem[{Chib and Ramamurthy(2010)}]{CR10}
Chib S, Ramamurthy S. 2010.
\newblock Tailored randomized block {MCMC} methods with application to {DSGE}
  models.
\newblock \emph{Journal of Econometrics} \textbf{155}: 19--38.

\bibitem[{Chotikapanich and Griffiths(2008)}]{CG08}
Chotikapanich D, Griffiths WE. 2008.
\newblock Estimating income distributions using a mixture of gamma densities.
\newblock In Chotikapanich D (ed.) \emph{Modeling Income Distributions and
  Lorenz Curves}. New York: Springer, 285--302.

\bibitem[{Clementi et~al.(2007)Clementi, Gallegati and Kaniadakis}]{CGK07}
Clementi F, Gallegati M, Kaniadakis G. 2007.
\newblock $\kappa$-generalized statistics in personal income distribution.
\newblock \emph{The European Physical Journal B} \textbf{57}: 187--193.

\bibitem[{Diebolt and Robert(1994)}]{DR94}
Diebolt J, Robert CP. 1994.
\newblock Estimation of finite mixture distributions through {Bayesian}
  sampling.
\newblock \emph{Journal of the Royal Statistical Society. Series B
  (Methodological)} \textbf{56}: 363--375.

\bibitem[{Doornik(2013)}]{D13}
Doornik JA. 2013.
\newblock \emph{Ox${}^{\text{TM}}$ 7: An Object-Oriented Matrix Programming
  Language}.
\newblock London: Timberlake Consultants Press.

\bibitem[{Dorfman(1979)}]{D79}
Dorfman R. 1979.
\newblock A formula for the {Gini} coefficient.
\newblock \emph{The Review of Economics and Statistics} \textbf{61}: 146--149.

\bibitem[{Eckernkemper and Gribisch(2021)}]{EB21}
Eckernkemper T, Gribisch B. 2021.
\newblock Classical and {Bayesian} inference for income distributions using
  grouped data.
\newblock \emph{Oxford Bulletin of Economics and Statistics} \textbf{83}:
  32--65.

\bibitem[{Escobar and West(1995)}]{EW95}
Escobar MD, West M. 1995.
\newblock Bayesian density estimation and inference using mixtures.
\newblock \emph{Journal of the American Statistical Association} \textbf{90}:
  577--588.

\bibitem[{Gastwirth(1972)}]{G72}
Gastwirth JL. 1972.
\newblock The estimation of the {Lorenz} curve and {Gini} index.
\newblock \emph{The Review of Economics and Statistics} \textbf{54}: 306--316.

\bibitem[{Gau et~al.(2014)Gau, {de Dieu\ Tapsoba} and Lee}]{GdL14}
Gau SL, {de Dieu\ Tapsoba} J, Lee SM. 2014.
\newblock {Bayesian} approach for mixture models with grouped data.
\newblock \emph{Computational Statistics} \textbf{29}: 1025--1043.

\bibitem[{Green(1995)}]{G95}
Green PJ. 1995.
\newblock Reversible jump {Markov} chain {Monte} {Carlo} computation and
  {Bayesian} model determination.
\newblock \emph{Biometrika} \textbf{82}: 711--732.

\bibitem[{Griffiths et~al.(2005)Griffiths, Chotikapanich and Rao}]{GCR05}
Griffiths WE, Chotikapanich D, Rao DSP. 2005.
\newblock Averaging income distributions.
\newblock \emph{Bulletin of Economic Research} \textbf{57}: 347 -- 367.

\bibitem[{Griffiths and Hajargasht(2012)}]{GH12}
Griffiths WE, Hajargasht G. 2012.
\newblock {GMM} estimation of mixtures from grouped data: An application to
  income distribution.
\newblock Department of Economics - Working Papers Series 1148, The University
  of Melbourne.

\bibitem[{Kakamu and Nishino(2019)}]{KN19}
Kakamu K, Nishino H. 2019.
\newblock Bayesian estimation of beta-type distribution parameters based on
  grouped data.
\newblock \emph{Computational Economics} \textbf{54}: 625--645.

\bibitem[{{Kom\`arek}(2009)}]{K09}
{Kom\`arek} A. 2009.
\newblock A new {R} package for {Bayesian} estimation of multivariate normal
  mixtures allowing for selection of the number of components and
  interval-censored data.
\newblock \emph{Computational Statistics \& Data Analysis} \textbf{53}:
  3932--3947.

\bibitem[{Lubrano and Ndoye(2016)}]{LN16}
Lubrano M, Ndoye AAJ. 2016.
\newblock Income inequality decomposition using a finite mixture of log-normal
  distributions: A {Bayesian} approach.
\newblock \emph{Computational Statistics \& Data Analysis} \textbf{100}:
  830--846.

\bibitem[{Malsiner-Walli et~al.(2016)Malsiner-Walli, Fr{\"u}hwirth-Schnatter
  and Gr{\"u}n}]{MFG16}
Malsiner-Walli G, Fr{\"u}hwirth-Schnatter S, Gr{\"u}n B. 2016.
\newblock Model-based clustering based on sparse finite {Gaussian} mixtures.
\newblock \emph{Statistics and Computing} \textbf{26}: 303--324.

\bibitem[{McDonald(1984)}]{M84}
McDonald JB. 1984.
\newblock Some generalized functions for the size distribution of income.
\newblock \emph{Econometrica} \textbf{52}: 647--663.

\bibitem[{McDonald and Ransom(1979)}]{MR79}
McDonald JB, Ransom MR. 1979.
\newblock Functional forms, estimation techniques and the distribution of
  income.
\newblock \emph{Econometrica} \textbf{47}: 1513--1525.

\bibitem[{McDonald and Xu(1995)}]{MX95}
McDonald JB, Xu YJ. 1995.
\newblock A generalization of the beta distribution with applications.
\newblock \emph{Journal of Econometrics} \textbf{66}: 133--152.

\bibitem[{Miller and Harrison(2013)}]{MH13}
Miller JW, Harrison MT. 2013.
\newblock A simple example of {Dirichlet} process mixture inconsistency for the
  number of components.
\newblock \emph{Advances in Neural Information Processing Systems} \textbf{26}:
  199--206.

\bibitem[{Newton and Raftery(1994)}]{NR94}
Newton MA, Raftery AE. 1994.
\newblock Approximate {Bayesian} inference with the weighted likelihood
  bootstrap (with discussion).
\newblock \emph{Journal of the Royal Statistical Society. Series B
  (Methodological)} \textbf{56}: 3--48.

\bibitem[{Nishino and Kakamu(2011)}]{NK11}
Nishino H, Kakamu K. 2011.
\newblock Grouped data estimation and testing of {Gini} coefficients using
  lognormal distributions.
\newblock \emph{Sankhy{\=a}: The Indian Journal of Statistics, Series B}
  \textbf{73}: 193--210.

\bibitem[{Paap and {van Dijk}(1998)}]{PvD98}
Paap R, {van Dijk} HK. 1998.
\newblock Distribution and mobility of wealth of nations.
\newblock \emph{European Economic Review} \textbf{42}: 1269--1293.

\bibitem[{Papastamoulis and Iliopoulos(2009)}]{PI09}
Papastamoulis P, Iliopoulos G. 2009.
\newblock Reversible jump {MCMC} in mixtures of normal distributions with the
  same component means.
\newblock \emph{Computational Statistics \& Data Analysis} \textbf{53}:
  900--911.

\bibitem[{{R Core Team}(2025)}]{R25}
{R Core Team}. 2025.
\newblock \emph{R: A Language and Environment for Statistical Computing}.
\newblock R Foundation for Statistical Computing, Vienna, Austria.

\bibitem[{Reed and Jorgensen(2004)}]{RJ04}
Reed WJ, Jorgensen M. 2004.
\newblock The double {Pareto}-lognormal distribution --- {A} new parametric
  model for size distributions.
\newblock \emph{Communications in Statistics - Theory and Methods} \textbf{33}:
  1733--1753.

\bibitem[{Richardson and Green(1997)}]{RG97}
Richardson S, Green PJ. 1997.
\newblock On {Bayesian} analysis of mixtures with an unknown number of
  components (with discussion).
\newblock \emph{Journal of the Royal Statistical Society: Series B (Statistical
  Methodology)} \textbf{59}: 731--792.

\bibitem[{Richardson and Green(1998)}]{RG98}
Richardson S, Green PJ. 1998.
\newblock Corrigendum: On {Bayesian} analysis of mixtures with an unknown
  number of components.
\newblock \emph{Journal of the Royal Statistical Society: Series B (Statistical
  Methodology)} \textbf{60}: 661--661.

\bibitem[{Robertson and Cryer(1974)}]{RC74}
Robertson T, Cryer JD. 1974.
\newblock An iterative procedure for estimating the mode.
\newblock \emph{Journal of the American Statistical Association} \textbf{69}:
  1012--1016.

\bibitem[{Spiegelhalter et~al.(2002)Spiegelhalter, Best, Carlin and {van der
  Linde}}]{SBCvdL02}
Spiegelhalter DJ, Best NG, Carlin BP, {van der Linde} A. 2002.
\newblock Bayesian measures of model complexity and fit.
\newblock \emph{Journal of the Royal Statistical Society: Series B (Statistical
  Methodology)} \textbf{64}: 583--639.

\bibitem[{Wiper et~al.(2001)Wiper, {Rios Insua} and Ruggeri}]{WIR01}
Wiper M, {Rios Insua} D, Ruggeri F. 2001.
\newblock Mixtures of gamma distributions with applications.
\newblock \emph{Journal of Computational and Graphical Statistics} \textbf{10}:
  440--454.

\end{thebibliography}

\clearpage
\renewcommand{\baselinestretch}{1.0}

\clearpage
\renewcommand{\baselinestretch}{1.0}
\begin{table}
	\caption{Simulated data 1: The conditional posterior estimates on $R=3$}
	\label{tab:sim}
	\begin{center}
		\begin{tabular}{crrrrrrl}
			\hline
			& \multicolumn{1}{c}{TRUE} & \multicolumn{1}{c}{MEAN} & \multicolumn{1}{c}{SD} & \multicolumn{4}{c}{95\%CI}\\
			\hline
			$\pi_{1}$ & 0.2 & 0.375 & 0.148 & [ & 0.152 & 0.600 & ]\\
			$\pi_{2}$ & 0.5 & 0.423 & 0.094 & [ & 0.301 & 0.617 & ]\\
			$\pi_{3}$ & 0.3 & 0.202 & 0.082 & [ & 0.068 & 0.339 & ]\\
			$\mu_{1}$ & 2.0 & 2.498 & 0.414 & [ & 1.854 & 3.007 & ]\\
			$\mu_{2}$ & 3.0 & 3.007 & 0.019 & [ & 2.966 & 3.044 & ]\\
			$\mu_{3}$ & 4.0 & 4.027 & 0.062 & [ & 3.887 & 4.128 & ]\\
			$\sigma_{1}^{2}$ & 0.3 & 0.703 & 0.409 & [ & 0.051 & 1.275 & ]\\
			$\sigma_{2}^{2}$ & 0.1 & 0.139 & 0.217 & [ & 0.070 & 1.173 & ]\\
			$\sigma_{3}^{2}$ & 0.2 & 0.153 & 0.062 & [ & 0.048 & 0.283 & ]\\
			\hline
		\end{tabular}
		\begin{minipage}{.6\textwidth}
			Note:
			Posterior means (MEAN), standard deviations (SD), and 95\% credible intervals (95\%CI) are displayed.
		\end{minipage}
	\end{center}
\end{table}

\begin{table}
	\caption{Simulated data 1: Marginal likelihoods}
	\label{tab:sim1-ML}
	\begin{center}
		\begin{tabular}{lrr}
			\hline
			& log ML & SE\\
			\hline
			MLN   &    2.843 & 0.478\\
			$R=3$ &   -1.342 & 0.146\\
			GB2   &  -76.923 & 0.307\\
			\hline
		\end{tabular}

		\begin{minipage}{.6\textwidth}
			Note:
			Log of marginal likelihoods (log ML) and the standard errors (SE), which are calculated using delta method, are displayed.
		\end{minipage}
	\end{center}
\end{table}

\begin{table}
	\caption{Simulated data 1: The posterior estimates of the Gini coefficients}
	\label{tab:sim1-gini}
	\begin{center}
		\begin{tabular}{lrrrrrrl}
			\hline
				& MEAN & MODE & SD & \multicolumn{4}{c}{95\%CI}\\
			\hline
			MLN   & 0.437 & 0.422 & 0.022 & [ & 0.410 & 0.487 & ]\\
			$R=3$ & 0.437 & 0.422 & 0.020 & [ & 0.413 & 0.485 & ]\\
			GB2   & 0.516 & 0.516 & 0.021 & [ & 0.465 & 0.552 & ]\\
			\hline
		\end{tabular}		
		\begin{minipage}{.6\textwidth}
			Note:
			Posterior means (MEAN), posterior modes (MODE), standard deviations (SD), and 95\% credible intervals (95\%CI) are displayed.
		\end{minipage}
	\end{center}
\end{table}

\begin{table}
	\caption{Simulated data 2: Marginal likelihoods}
	\label{tab:sim2-ML}
	\begin{center}
		\begin{tabular}{lrr}
			\hline
			& log ML & SE\\
			\hline
			MLN   &   9.977 & 0.473\\
			$R=2$ &   8.338 & 0.277\\
			GB2   &  10.743 & 0.115\\
			\hline
		\end{tabular}

		\begin{minipage}{.6\textwidth}
			Note:
			Log of marginal likelihoods (log ML) and the standard errors (SE), which are calculated using delta method, are displayed.
		\end{minipage}
	\end{center}
\end{table}

\begin{table}
	\caption{Simulated data 2: The posterior estimates of the Gini coefficients}
	\label{tab:sim2-gini}
	\begin{center}
		\begin{tabular}{lrrrrrrl}
			\hline
				& MEAN & MODE & SD & \multicolumn{4}{c}{95\%CI}\\
			\hline
			MLN   & 0.333 & 0.330 & 0.010 & [ & 0.321 & 0.356 & ]\\
			$R=2$ & 0.336 & 0.332 & 0.010 & [ & 0.322 & 0.360 & ]\\
			GB2   & 0.342 & 0.339 & 0.006 & [ & 0.331 & 0.354 & ]\\
			\hline
		\end{tabular}		
		\begin{minipage}{.6\textwidth}
			Note:
			Posterior means (MEAN), posterior modes (MODE), standard deviations (SD), and 95\% credible intervals (95\%CI) are displayed.
		\end{minipage}
	\end{center}
\end{table}

\begin{table}
	\caption{Marginal likelihoods for FIES data in 2020}
	\label{tab:emp-ml}
	\begin{center}
		\begin{tabular}{lrrcrr}
			\hline
			& \multicolumn{2}{c}{Two-or-more person} && \multicolumn{2}{c}{Workers'}\\
			\cline{2-3}\cline{5-6}
			& log ML & SE && log ML & SE\\
			\hline
			MLN   &  14.850 & 0.725 && 17.888 & 0.229\\
			$R=2$ &  16.103 & 0.249 && 11.920 & 0.302\\
			GB2   & -20.798 & 0.230 && 10.685 & 0.257\\
			\hline
		\end{tabular}		
		\begin{minipage}{.6\textwidth}
			Note:
			Log of marginal likelihoods (log ML) and the standard errors (SE), which are calculated using delta method, are displayed.
		\end{minipage}
	\end{center}
\end{table}

\begin{table}
	\caption{The conditional posterior estimates from the FIES data in 2020}
	\label{tab:emp}
	\begin{center}
		\begin{tabular}{crrrrrlcrrrrrl}
			\hline
			& \multicolumn{6}{c}{Two-or-more person} && \multicolumn{6}{c}{Workers'}\\
			\cline{2-7}\cline{9-14}
			& \multicolumn{1}{c}{MEAN} & \multicolumn{1}{c}{SD} & \multicolumn{4}{c}{95\%CI} && \multicolumn{1}{c}{MEAN} & \multicolumn{1}{c}{SD} & \multicolumn{4}{c}{95\%CI}\\
			\hline
			$\pi_{1}$ & 0.355 & 0.107 & [ & 0.173 & 0.592 & ] && 0.674 & 0.211 & [ & 0.133 & 0.944 & ]\\
			$\pi_{2}$ & 0.645 & 0.107 & [ & 0.408 & 0.827 & ] && 0.326 & 0.211 & [ & 0.056 & 0.867 & ]\\
			$\mu_{1}$ & 1.221 & 0.070 & [ & 1.111 & 1.384 & ] && 1.793 & 0.098 & [ & 1.515 & 1.893 & ]\\
			$\mu_{2}$ & 1.951 & 0.080 & [ & 1.813 & 2.128 & ] && 2.114 & 0.079 & [ & 1.942 & 2.249 & ]\\
			$\sigma_{1}^{2}$ & 0.116 & 0.024 & [ & 0.074 & 0.169 & ] && 0.202 & 0.053 & [ & 0.130 & 0.271 & ]\\
			$\sigma_{2}^{2}$ & 0.195 & 0.031 & [ & 0.134 & 0.255 & ] && 0.116 & 0.068 & [ & 0.046 & 0.179 & ]\\
			\hline
		\end{tabular}		
		\begin{minipage}{.7\textwidth}
			Note:
			Posterior means (MEAN), standard deviations (SD), and 95\% credible intervals (95\%CI) are displayed.
		\end{minipage}
	\end{center}
\end{table}

\begin{table}
	\caption{The posterior estimates of the Gini coefficients from the FIES data in 2020}
	\label{tab:emp-gini}
	\begin{center}
		\begin{tabular}{lrrrrrrlcrrrrrrl}
			\hline
			& \multicolumn{7}{c}{Two-or-more person} && \multicolumn{7}{c}{Workers'}\\
			\cline{2-8}\cline{10-16}
			& MEAN & MODE & SD & \multicolumn{4}{c}{95\%CI} && MEAN & MODE & SD & \multicolumn{4}{c}{95\%CI}\\
			\hline
			MLN   & 0.299 & 0.299 & 0.004 & [ & 0.294 & 0.306 & ] && 0.245 & 0.244 & 0.006 & [ & 0.239 & 0.251 & ]\\
			$R=2$ & 0.299 & 0.299 & 0.003 & [ & 0.294 & 0.305 & ] && 0.244 & 0.244 & 0.003 & [ & 0.239 & 0.250 & ]\\
			GB2   & 0.316 & 0.316 & 0.004 & [ & 0.309 & 0.324 & ] && 0.249 & 0.249 & 0.003 & [ & 0.244 & 0.254 & ]\\
			\hline
		\end{tabular}
		\begin{minipage}{.9\textwidth}
			Note:
			Posterior means (MEAN), posterior modes (MODE), standard deviations (SD), and 95\% credible intervals (95\%CI) are displayed.
		\end{minipage}
	\end{center}
\end{table}
\end{document}